\documentclass[usenatbib]{mn2e}

\usepackage{psfig,url}

\footnotesize
\newdimen\digitwidth    
\setbox0=\hbox{\rm0}
\digitwidth=\wd0
\catcode`!=\active
\def!{\kern\digitwidth}
\normalsize

\title[Isolated pulsar evolution]
{Isolated pulsar spin evolution on the $P$--$\dot{P}$ Diagram}
\author[J.P.~Ridley \& D.R.~Lorimer]
{J.P.~Ridley$^{1}$ and D.R.~Lorimer$^{1,2}$
\\
$^1$Department of Physics, West Virginia University, PO~Box~6315, Morgantown,
WV~26506, USA\\
$^2$National Radio Astronomy Observatory, PO Box 2, Green Bank, WV~24944, USA\\
}
\let\leqslant=\leq 
\date{Accepted 2010 January 13. Received 2009 December 21; in original form 2009 July 15}
\begin{document}
\maketitle
\newcommand{\setthebls}{
}

\setthebls

\begin{abstract} 
We look at two contrasting spin-down models for isolated radio pulsars
and, accounting for selection effects, synthesize observable
populations. While our goal is to reproduce all of the observable
characteristics, in this paper we pay particular attention to the form
of the spin period vs. period derivative ($P$--$\dot{P}$) diagram and
its dependence on various pulsar properties.  We analyse the initial
spin period, the braking index, the magnetic field, various beaming
models, as well as the pulsar's luminosity.  In addition to
considering the standard magnetic dipole model for pulsar spin-down,
we also consider the recent hybrid model proposed by Contopoulos \&
Spitkovsky.  The magnetic dipole model, however, does a better job of
reproducing the observed pulsar population.  We conclude that random
alignment angles and period dependent luminosity distributions are
essential to reproduce the observed $P$--$\dot{P}$ diagram. We also
consider the time decay of alignment angles, and attempt to reconcile
various models currently being studied.  We conclude that, in order to
account for recent evidence for the alignment found by Weltevrede \&
Johnston, the braking torque on a neutron star should not depend
strongly on the inclination. Our simulation code is publically
available and includes a web-based interface to examine the results
and make predictions for yields of current and future surveys.
\end{abstract}


\section{INTRODUCTION}\label{sec:intro}

The statistical properties of the radio pulsar population provide
valuable constraints on the birth properties, evolution and radio
lifetimes of neutron stars and have been the subject of many studies
over the years \citep[see,
e.g.,][]{go70,tm77,lmt85,no90,bwhv92,joh94,lbh97,hbwv97,acc02,gob+02,fk06}.
A key diagram to reproduce is the logarithmic plot of pulsar periods,
$P$, and their rates of slowdown, $\dot{P}$, usually referred to as
the $P$--$\dot{P}$ diagram.  The many factors that affect the
distribution of pulsars on the $P$--$\dot{P}$ diagram include
the evolution of pulsars' spin-down, luminosity, age, magnetic field,
beaming, and the angle between the magnetic field and the rotation
axis.  While each parameter individually affects how pulsars evolve
across the diagram, covariances between multiple parameters also
influence the results.

The goal of this paper is to compare two pulsar spin-down models using
Monte Carlo simulations of the Galactic young pulsar population. The first
model we consider treats the pulsar as a rotating magnetic dipole from
which there is a simple relationship between the braking torque and
the magnetic field \citep{gol68,pac68}. This model has been used
extensively in pulsar population studies in the past \citep[see,
e.g.][]{lmt85,no90,bwhv92} and has most recently been implemented by
\citet{fk06}, hereafter FK06.  In this paper, we desire to undergo a 
direct comparison of their results with our simulations.

The other model we consider, by \citet{cs06}, hereafter CS06, treats the 
pulsar as a combination of a misaligned magnetic dipole rotating in vacuum 
and an aligned magnetic dipole rotating in an atmosphere with ideal 
magnetohydrodynamic conditions.  In addition to implementing these two 
models with parameters as described in detail in FK06 and CS06, we 
investigate the effects of changing one or more of the assumptions made 
by these authors in an attempt to obtain a more accurate interpretation of 
the relationships between these parameters and the resulting
$P$--$\dot{P}$ diagram.

As described in Section \ref{sec:code}, we have created a computer
simulation package which implements the above models and allows us to
simulate the distribution of pulsars in the Galaxy and change the
individual properties of each one.  By ``observing'' this population
with accurate models of completed pulsar surveys (Section
\ref{sec:code2}), we are able to create a simulated sample and compare
its properties to those of the real observed sample. In Section
\ref{sec:comparison} we briefly describe our method for comparing two
population samples.  In Section \ref{sec:results}, we present and
discuss our results. Our main conclusions are summarized in Section
\ref{sec:conclusions}. To facilitate comparison with other studies,
and provide a tool for the community, a major feature of this work is
to make our simulation code freely available on a website which is
discussed in Appendix A.

\section{Modeling the underlying pulsar population}\label{sec:code}

The computer program we have developed for this work, {\tt evolve}, is
an extension of the software package, {\tt psrpop} which was initially
developed to study the Galactic distribution of pulsars detected in
the Parkes Multibeam surveys \citep{lfl+06}. For that work,
time-independent pulsar populations were synthesized and compared to
the observed sample. The main aim of this paper is to develop a
time-dependent model of the population and is realized through a
program we call {\tt evolve}. In this section, we describe the logical
flow of the program which determines first whether a model pulsar
is potentially observable (i.e.~is radio loud and beaming towards us),
before computing any kinematic or spatial parameters.

The first step is to randomly assign an age, $t$, and a magnetic
field, $B$, to each pulsar.  The age is selected between 0 and $t_{\rm
  max}$, with $t_{\rm max}$ being $10^9$ years.  As per FK06, the magnetic 
field is determined by a log-normal distribution, centred around a mean value,
$\mu_{\log B}=12.65$, with a standard deviation $\sigma_{\log B}=0.55$. 
The initial period, $P_0$, is then randomly generated.  Following
FK06, $P_0$ is chosen from a Gaussian distribution 
centred around $\mu_P=300$ ms with a standard deviation $\sigma_P=150$ ms. 

\begin{figure}
\psfig{file=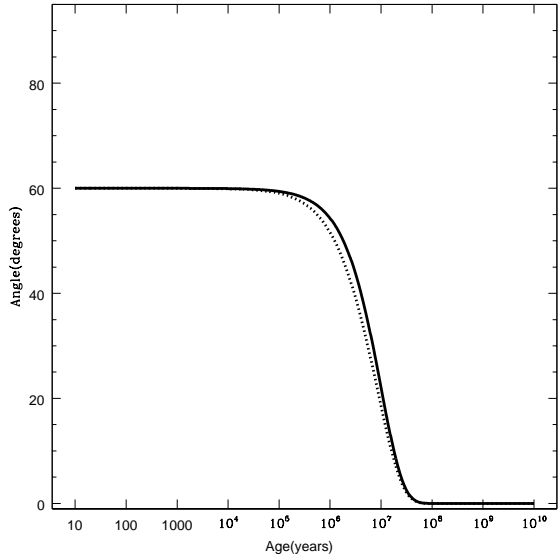,width=8cm,angle=0}
\caption{The time dependence of the alignment angle shown for 
  Equation~\ref{eq:chi} (dotted curve) and the model
  used in WJ08 (solid curve).  We use $60^{\circ}$ for a reference as roughly 
half of all angles will be above and half will be below that value for a 
given random distribution of angles.}  
\label{fig:compare}
\end{figure}

We next compute the inclination angle between the magnetic field axis and
spin axis, $\chi$. In implementations of the magnetic dipole model,
it is usually assumed (see, e.g., FK06) that $\chi=90^{\circ}$.  However, 
a more realistic model would account for a distribution of $\chi$ and use 
this information in a self-consistent way when calculating the beaming 
fraction discussed below. We explore a number of options, first fixing
$\chi=90^{\circ}$, and subsequently considering a random distribution
of angles by selecting $\cos{\chi}$ from a flat distribution
between 0 and 1. The latter choice effectively orients the magnetic
axis as a uniform random vector over $4\pi$~sr.  Finally, following recent 
empirical evidence in favour of a secular decay of $\chi$ with time 
\citep[][ hereafter WJ08]{wj08}, we also consider an exponentially decaying 
model defined as
\begin{equation}
\sin{\chi}=\sin{\chi_0}\exp(-t/t_d),
\label{eq:chi}
\end{equation}
where $\chi_0$ is the initial angle (which we again choose by
selecting $\cos \chi_0$ from a 0--1 flat distribution), and $t_d$ is
the $1/e$ decay timescale.  Note that we choose this decay law over
the one considered by WJ08 for mathematical and computational
convenience. As shown in Fig.~\ref{fig:compare}, the two dependencies
give similar results.
The braking index, $n$, is then assigned.  In general, $n$ is related to
$\nu$, the frequency of rotation, $\nu=1/P$, and its time derivative
\begin{equation}
\dot{\nu} = -K \nu^n,
\label{eq:nu}
\end{equation}
where $K$ is a constant that depends on the pulsar's moment of inertia and
radius \citep[see, e.g.][]{lk05}.  We can use a single index for all pulsars, 
or randomly assign an index to each pulsar. Specific choices depend on the 
spin-down model under consideration which are described below.

Each spin-down model yields a time-dependent equation for the pulsar's
period and period derivative.  The pulsar's period evolves in time
depending on the spin-down model, the magnetic field, the braking
index, and the alignment angle.  Considering first the magnetic dipole
model, the period dependence can be found by solving
\begin{equation}
P^{n-2}\dot{P}=kB^2\sin^2\chi(t),
\label{eq:periodstart}
\end{equation}
where $\sin\chi(t)$ is defined by Equation \ref{eq:chi}.  The constant 
\begin{equation}
k =\frac{8\pi^2R^6}{3Ic^3}, 
\end{equation}
where $R$ is the radius of the neutron star, assumed to be 10~km, and
$I$ is the moment of inertia, assumed to be $10^{38}$~kg~m$^2$.  For
this choice of parameters, $k$ takes the numerical value $9.768 \times
10^{-40}$, $B$ is in units of Gauss while $P$, $P_0$ and $t$ are in
seconds.

The implementation of this model in FK06 assumes $n=3$ and 
$\chi=90^{\circ}$.  In this case, the period dependence
\begin{equation}
P(t) = \sqrt{P_0^2+2kB^2t}.
\label{eq:pfk06a}
\end{equation}
More generally, for any braking index other than $n=1$, 
and a constant value of $\chi$, we find
\begin{equation}
P(t) = [P_0^{n-1}+(n-1)kB^2t\sin^2\chi]^{\frac{1}{n-1}}.
\label{eq:pfk06b}
\end{equation}
Finally, extending this result for a time dependent inclination
angle of the form assumed in Equation~\ref{eq:chi}, we find
\begin{equation}
P(t) = \left[P_0^{n-1}+\frac{(n-1)}{2}t_dkB^2\sin^2 \chi_0 
\left(1-e^{-2t/t_d}\right)\right]^{\frac{1}{n-1}}.
\label{eq:pfk06c}
\end{equation}

Depending on the model under consideration, one of
Equations~\ref{eq:pfk06a}--\ref{eq:pfk06c} is used to calculate the
pulsar's current period, $P$.  Equation~\ref{eq:periodstart} is then
used to find the corresponding value of $\dot{P}$. At this point, a
simple test is made to determine if the pulsar has crossed the
so-called death line.  The death line signifies when a pulsar becomes
radio-quiet, and \citet{bwhv92} quantified it as the locus of points
for which
\begin{equation}
\frac{B}{P^2} = 0.17\times10^{12} \textrm{ G s}^{-2}.
\label{eq:deathline}
\end{equation}
If the pulsar has crossed the death line, the simulation marks the
pulsar as dead (i.e.~radio quiet) and moves on to create the next
pulsar.  If the pulsar has not crossed the line, then we continue on
with the simulation.  It should be noted, however, that a small number 
of pulsars \citep[e.g. J2144$-$3933;][]{ymj99} have crossed this 
theoretical line and are still observable at radio frequencies.

For our implementation of the CS06 model, the technique is slightly
different.  The time evolved period and period-derivative are
calculated by integrating and then numerically solving
\begin{eqnarray}
\lefteqn{\dot{P}=3.3\times
  10^{-16}\left(\frac{P}{P_0}\right)^{2-n}\left(\frac{B}{10^{12}\textrm{ G}}\right)^2\left(\frac{P_0}{1
\textrm{ s}}\right)^{-1}}\nonumber\\
& & \left(1-\frac{P}{P_{\rm death}}\cos^2\chi\right),
\label{eq:pcs06}
\end{eqnarray}
where 
\begin{equation}
P_{\rm death}=\left[0.81\times\left(\frac{B}{10^{12}\textrm{ G}}\right)\left(\frac{1\textrm{ s}}{P_0}\right)\right]^{\frac{2}{n+1}}\textrm{ s}.
\label{eq:pdeath}
\end{equation}
In this case, the pulsar is declared dead when the period becomes
greater than $P_{\rm death}$.

To account for the fact the pulsar radiation is only beamed to some
fraction of $4\pi$~sr, a variety of beaming models have been
implemented in {\tt evolve}.  The default model is from the work of
\cite{tm98}, henceforth TM98, who found an empirical relationship for
the fraction of the whole sky illuminated by a pulsar:
\begin{equation}
f(P) = 0.09\left[\log\left(\frac{P}{1\textrm{
s}}\right)-1\right]^2+0.03.
\label{eq:tm98beaming}
\end{equation}
Other models from \cite{big90b}, \cite{lm88}, and \cite{nv83},
henceforth B90, LM88, and NV83 respectively, determine the beaming
fraction from the angular beam radius
\begin{equation}
\rho = \rho_0 P^{\gamma},
\label{eq:beam}
\end{equation}
where the constants $\rho_0$ and $\gamma$ are uniquely defined in each
model.  Assuming that radio waves are emitted from both poles of the 
pulsar and that the beams are circular beams, and following TM98 leads 
to a beaming fraction dependence on magnetic inclination angle of
\begin{equation}
f(\rho,\chi) = \left\{ 
\begin{array}{l l l l}
  2\sin{\chi}\sin{\rho}& \quad \chi>\rho, \chi+\rho < \frac{\pi}{2}\\
  \cos{(\chi-\rho)} & \quad \chi>\rho, \chi+\rho > \frac{\pi}{2}\\ 
  1-\cos{(\chi+\rho)} & \quad \chi<\rho, \chi+\rho < \frac{\pi}{2}\\
  1 & \quad \chi<\rho, \chi+\rho > \frac{\pi}{2}.\\
\end{array} \right.
\label{eq:fpchi}
\end{equation}
Also available is a simple period-independent beaming model that
assumes $f=0.2$ for all values of $P$. To determine whether a pulsar
is beaming towards us or not, we compare the value of $f$ computed in
one of the above ways with a random number between 0 and 1. Only those
pulsars for which this random number is less than or equal to $f$ are
deemed to be beaming towards us and we move on to the calculations of
the final properties of the pulsar.

To describe the location of model pulsars in the Galaxy, we use a
regular Cartesian ($x,y,z$) coordinate system with the Galactic center
at the origin and the position of the Sun at ($0,8.5,0$) kpc, again 
identical to the procedure performed by FK06.  Each
pulsar's initial $x$ and $y$ positions on the Galactic plane are
determined by randomly choosing a point in the plane of our Galaxy
along the spiral arms.  We follow the procedure described by FK06 to
implement the spiral arm structure.  For the radial distribution, we
use the form described by \cite{yk04}.  Then, dispersion away from the
plane is used to determine the initial $z$ position.  For this, we use
an exponential distribution with a mean $z_0$=50 pc, and then
arbitrarily choose a sign.

To model pulsar birth velocities, we use the optimal model found in
FK06, in which the initial velocity components of each pulsar in the
$x$, $y$ and $z$ directions are drawn from exponential distributions
with a mean absolute value of 180~km~s$^{-1}$.  Although other distributions are
available within {\tt evolve}, we do not investigate them any further
here. To model the resulting evolution in the Galactic gravitational
potential, we follow the method described by \citet{lbdh93}, where the
integration in the gravitational galactic potential is done using a
model described by \cite{ci87}.  From the final position in the model
galaxy, we compute the pulsar's distance from the Sun, $d$, and its
expected dispersion measure (DM) and scatter-broadening timescale at
1~GHz.  The latter two quantities are derived using the NE2001 model
for the Galactic distribution of free electrons~\citep{cl02a}.

The radio luminosity is calculated, following FK06, with a $P$ and $\dot{P}$
power law dependence, defined by
\begin{equation}
\log{L} = \log{L_0} + \alpha \log{P} + \beta \log{(\dot{P}/10^{-15})}
          + \delta_L,
\label{eq:lum}
\end{equation}
where $L_0$ is 0.18 mJy kpc$^2$, and $\delta_L$ is randomly chosen
from a normal distribution with $\sigma_{\delta_L}$ = 0.8. The program
allows for $\alpha$ and $\beta$, two free parameters, to be varied to
determine the best functional dependence.

For comparison, we also make available in the code a simple model in
which the luminosity is independent of all other parameters.
Following FK06, we take the probability, $p(L)$, of a given luminosity
as being
\begin{equation}
p(L) \propto \left\{ 
\begin{array}{l l l}
  0                  & \quad L \in [0 \textrm{ mJy kpc$^2$, }0.1 \textrm{ mJy kpc$^2$})   \\
  L^{-\frac{19}{15}} & \quad L \in [0.1 \textrm{ mJy kpc$^2$, }2.0 \textrm{ mJy kpc$^2$}) \\ 
  L^{-2}             & \quad L \in [2.0 \textrm{ mJy kpc$^2$, }\infty).                   \\
\end{array} \right .
\label{eq:randlum}
\end{equation}

\section{Modeling the observed pulsar population}\label{sec:code2}

The steps described above allow us to create a population of synthetic
pulsars that are potentially observable. That is, they are deemed to
be beaming towards the Earth and the spin-down models we are
investigating predict that they are radio-loud. We wish to compare
this population with those pulsars that are potentially detectable by
current surveys. Since the primary objective of this work is to
reproduce and extend upon the work of FK06, in this paper, we focus on
the sample of pulsars detectable by the Parkes Multibeam
\citep[see][and references therein, hereafter PMB]{lfl+06} and
Swinburne \citep[][hereafter SMB]{ebvb01} multibeam pulsar surveys.
By accurately modeling the detection thresholds of these surveys as
described below, and selecting only those model pulsars that are
theoretically detectable, we can form samples of model observed
pulsars. When FK06 were carrying out their work, the sample of real
pulsars detected in these surveys was 1065. In our analysis, we will
use the sample of 1135 pulsars currently catalogued.  Our criteria is
to select all isolated pulsars not associated with a globular cluster and
having both $\dot{P}<10^{-12}$ and $P>30$ ms.

To form our model observed samples, the first step is to compute the
apparent flux density of each model pulsar, $S$. Following standard
practice \cite{lk05}, we neglect geometrical factors in the inverse
square law relationship and find $S$ from the pulsar's distance, $d$,
and radio luminosity, $L$, as follows
\begin{equation}
S = \frac{L}{d^2}.
\label{eq:flux}
\end{equation}
Note that this flux density is defined (following our definition of
$L$) to be at an observing frequency of 1.4~GHz, the frequency at
which the PMB and SMB surveys were carried out. For the purposes of
this work, no assumptions about the pulsar spectral index distribution
are necessary.

To model the detection threshold of the pulsar surveys, it is also
necessary to model the pulse widths. Following FK06, we assume that
the intrinsic pulse width, $W_{\rm int}$, is 5\% of the pulse period.
The observed pulse width, $W_{\rm obs}$, and signal-to-noise ratio,
S/N, are then calculated using the method described in \cite{lfl+06}
in their Equations 1--7.  Pulse width models from recent papers 
\citep[see, e.g.,][]{slk+09} have also been considered 
and are available for use in the code.

All pulsars which lie inside the sky boundaries covered by the two surveys, 
have S/N$>9$, and where $W_{\rm obs}<P$ are deemed detectable and saved 
for subsequent analysis, as described below.  Our simulation proceeds 
until 1135 pulsars are detected, to match the combined sample found in 
the PMB and SMB surveys.

\section{METHOD FOR COMPARISON}\label{sec:comparison}

To compare our trial results quantitatively, we use the
Kolmogorov-Smirnoff (KS) test, a non-parametric estimator based on the
maximum deviation seen in the cumulative distribution of two sample
data sets \citep[see, e.g.,][]{pftv86}.  For a given distribution, the
KS test returns a statistic that is used to determine a probability,
$Q$, that two samples came from the same underlying population. As
described below, we make use of this probability to investigate
whether model pulsar populations are inconsistent with the observed
data.
\begin{table} 
\caption{Here we show how the KS probability for the period distribution
  behaves when given simulated samples that are exactly the same as, 
  similar to, and completely different than an original sample.  See text 
  for details of the simulations.}
\label{tb:ks}
\begin{tabular}{cccc}
\hline
TRIAL & $Q_P$ & $Q_P$    & $Q_P$ \\
\#     & (SAME)& (SIMILAR) & (DIFFERENT) \\
\hline
1 & 0.1299 & 0.0296 & $< 10^{-12}$\\
2 & 0.8928 & 0.0094 & $< 10^{-12}$\\
3 & 0.7397 & 0.1098 & $< 10^{-12}$\\
4 & 0.6958 & 0.1744 & $< 10^{-12}$\\
\hline
AVG & 0.6145 & 0.0808 & $< 10^{-12}$\\
\hline
\end{tabular}
\end{table}

To investigate the sensitivity of the KS test to changing model
parameters, we run a single simulation using the optimal parameters
from the FK06 model.  We then generate four further Monte Carlo
realizations of the same model and compare them to the original
simulation using the KS probability for the period distribution,
$Q_P$. Next, we make a small modification, changing $\mu_P$ from 300
ms to 280 ms, run four more simulations and compute $Q_P$ by comparing
with our original simulation. Finally, we make a more significant
change to the model, changing $\mu_P$ to 250 ms, $\sigma_P$ to 100 ms,
and $\mu_{log B}$ to 12.00. We run four further simulations using
these parameters and record $Q_P$ against the original
simulation. From the results of these simulations that are summarized
in Table~\ref{tb:ks}, we conclude that individual KS probabilities
above 10\% suggest that two populations are consistent with another,
while probabilities in the range 1--10\% imply possible differences
between populations. However, the steep decline in the KS probability
when two populations differ significantly suggest that it is a
powerful metric to weigh one model against another.

In this work, we wish to track the multiple observed parameters
relevant to our simulations. Therefore, following \citet{bwhv92}, we
define a figure of merit for each simulation
\begin{equation}
{\rm FOM}
 = \log[Q_P\times Q_{\dot{P}}\times Q_l\times Q_b].
\label{eq:fom}
\end{equation}
Here the subscripts on each KS $Q$ value refer to the distributions of
period ($P$), period derivative ($\dot{P}$), Galactic longitude ($l$)
and Galactic latitude ($b$).

\section{RESULTS AND DISCUSSION}\label{sec:results}

\begin{figure*}
\psfig{file=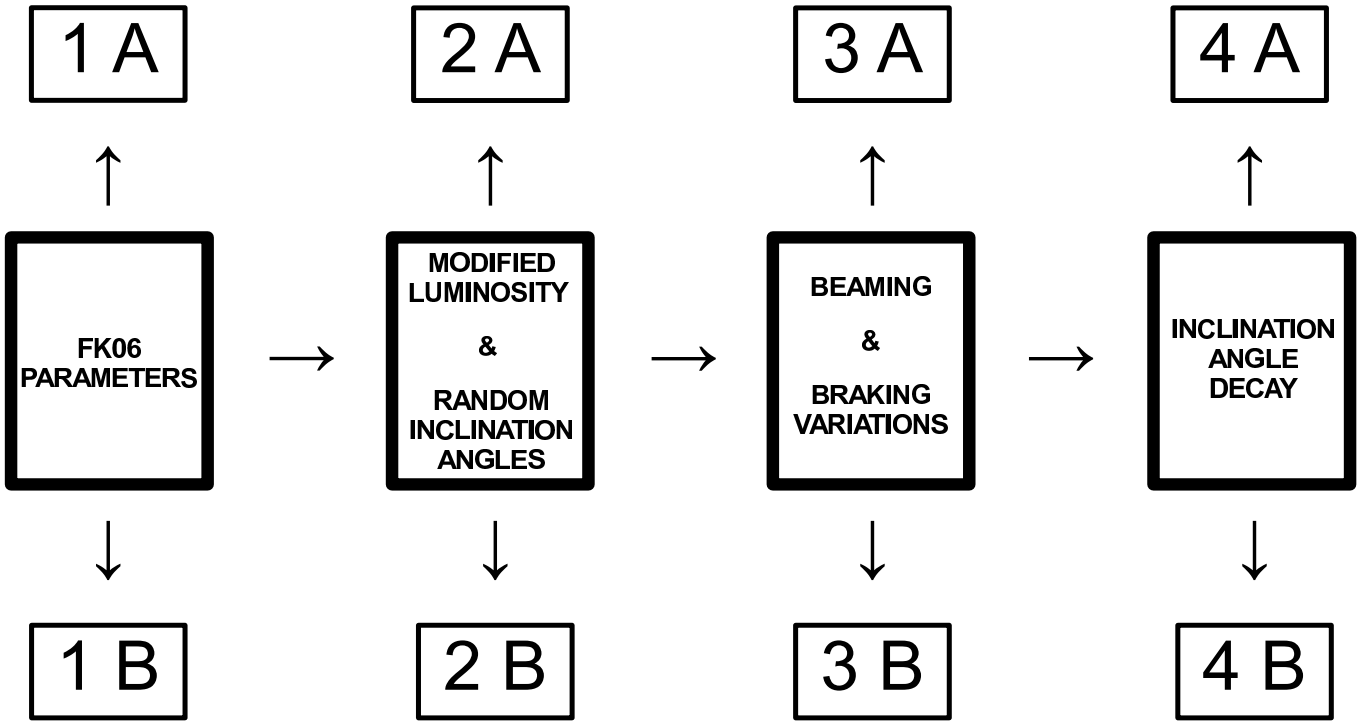,width=18cm,angle=0}
\caption{Schematic showing the logical flow of our approach to
comparing the power law (A) and CS06 spin-down (B) models. We begin in
model 1 by using the best parameters from FK06. As described in the
text, subsequent models
then contain additional complexity in an attempt to be more physically
plausible representations of the true population.}
\label{fig:flow}
\end{figure*}

In this section, we will describe and discuss our results. Our main
goal is to compare the power law spin-down model (Eq.~\ref{eq:nu})
with the new spin-down model proposed by CS06 (Eq.~\ref{eq:pcs06})
under a variety of assumptions. The logical flow of our approach is
summarized in Fig.~\ref{fig:flow}.

\subsection{Simulation Models}

We begin by using the basic model parameters from FK06 and applying
them to both spin-down laws which we call models 1A and 1B. In this
convention ``A'' refers to the power-law spin-down model and ``B''
refers to the CS06 spin-down model.  Model 2 attempts to improve the
original model by changing the luminosity law and using random
inclination angles to better match the observed sample. Subsequent
models are motivated to be more physically realistic with the
incorporation of: angle-dependent beaming models and a range of
braking indices for models 3A and 3B, and inclination angle decay for
models 4A and 4B.

Table \ref{tb:fom} provides an overall summary of our results, showing
base-10 logarithms of the individual KS probabilities and the FOM for
each model.  We also compute the mean birth rate (${\cal R}$) of
pulsars required to produce the observed sample sizes for each
model. This is defined as the ratio of the {\it total} number of
pulsars generated in each simulation to the maximum age of the
population, $t_{\rm max}$.

\subsubsection{Model 1: FK06 basic parameters}

\begin{figure}
\psfig{file=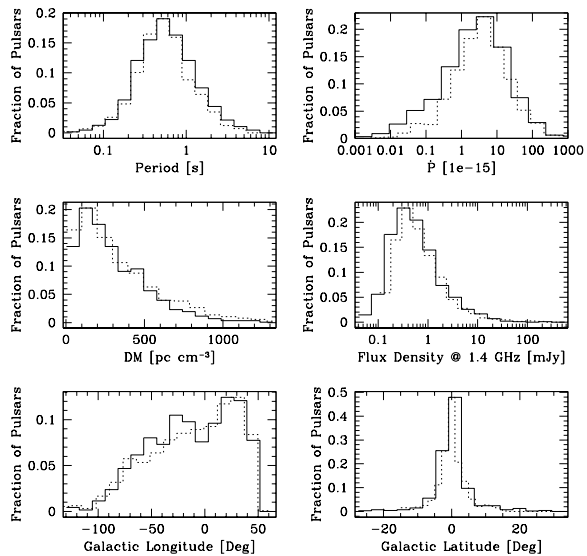,width=8cm,angle=0}
\caption{Histograms of some of the key properties from model 1A.  
The solid histograms represent the observed pulsar population.  The 
dotted histograms are the results of our simulation.}
\label{fig:model1}
\end{figure}

Adopting the optimal parameters found by FK06, we generate model A
which is shown in Fig.~\ref{fig:model1}.  The histograms show a
relatively accurate replication of the original FK06 results.  While
we are generally able to reproduce the results of FK06, we find
somewhat lower KS probabilities for $P$ and $\dot{P}$ than FK06.
Obtaining an exact match to results obtained with a different
simulation is challenging and the discrepancies we find highlight the
difference in the implementation of the model in the two codes.
We duplicate most of the major results well.  Some other parameters, 
such as $\alpha$ and $\beta$, have to be changed slightly from the values 
presented in FK06, but we obtain the majority of their results using 
our code.  These changes are addressed in the remaining models.

For model B, we retain all the parameters of model A with the
exception of the spin-down law, which we replace by our implementation
of CS06 described in Section~\ref{sec:code}.  As evidenced by the
results shown in Table~\ref{tb:fom}, the $P$ and $\dot{P}$
distributions are not as well reproduced.

This model has mixed results.  Many areas, such as the effect of the
alignment angle, the effect of having no angle dependence, and the
underlying distribution of pulsars lying below the death line (see
Fig.~1, 3, 6 of their paper), can be replicated with our code.
However, the KS statistics are not as high as they are for the FK06
spin-down model.  This applies not only to Model 1, but to the rest of
the models as well.

\begin{figure}
\psfig{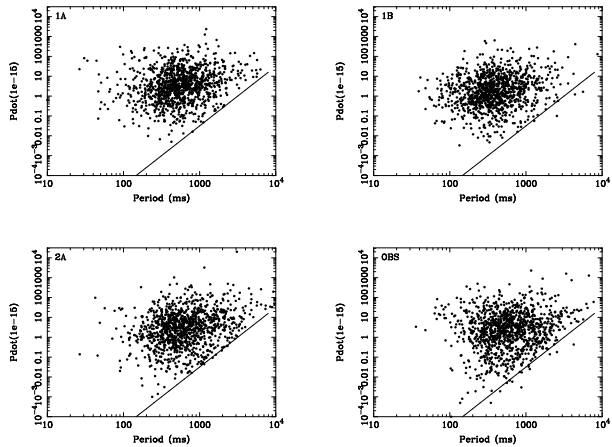}
\caption{
  Our resulting $P$--$\dot{P}$ diagrams, with the adopted death line
  shown in each plot for reference. Shown in the upper panels are our
  implementations of FK06 (model 1A, upper left) and CS06 (model 1B,
  upper right). The lower panels show our improved simulation (model
  2A, lower left), which is based on an FK06 spin-down model, and the
  real observed sample of pulsars (lower right).
}
\label{fig:results}
\end{figure}

\begin{table*}
\caption{
  Summary of KS statistics, figures of merit (FOM) and
  birth rate ($\cal R$) comparing FK06 with the four models
  considered in this paper. See text for details of the various models.
  }
\label{tb:fom}
\begin{tabular}{l*{9}{r}}
\hline
               &         & \multicolumn{2}{c}{Model 1} & \multicolumn{2}{c}{Model 2} & \multicolumn{2}{c}{Model 3} & \multicolumn{2}{c}{Model 4}            \\
 Parameter     & FK06    & \multicolumn{1}{c}{A} & \multicolumn{1}{c}{B} & \multicolumn{1}{c}{A} & \multicolumn{1}{c}{B} & \multicolumn{1}{c}{A} & \multicolumn{1}{c}{B} & \multicolumn{1}{c}{A} & \multicolumn{1}{c}{B}       \\
\hline
$P$            & $-$2.15 & $-$1.68 & $-$12.0  & $-$0.94 & $-$7.00  & $-$1.75 & $-$12.0 & $-$12.0 & $-$10.7 \\
$\dot{P}$      & $-$1.20 & $-$4.41 & $-$5.35  & $-$0.51 & $-$0.67  & $-$0.08 & $-$1.76 & $-$6.37 & $-$2.21 \\
$l$            & $-$0.90 & $-$0.12 & $-$0.12  & $-$0.07 & $-$0.09  & $-$0.28 & $-$0.40 & $-$0.10 & $-$0.25 \\
$b$            & $-$1.14 & $-$0.50 & $-$0.13  & $-$0.23 & $-$0.30  & $-$0.95 & $-$1.87 & $-$1.01 & $-$1.56 \\
\hline
FOM            & $-$5.39 & $-$6.71 & $-$17.6  & $-$1.75 & $-$8.06  & $-$3.06 & $-$16.0 & $-$19.5 & $-$14.7 \\
${\cal R}$ 
(psrs/century) & 2.8     & 3.85    & 1.97     & 3.56    & 2.84     & 2.97    & 3.75    & 5.31    & 3.68    \\
\hline
\end{tabular}
\end{table*}

\subsubsection{Model 2: Modified luminosity law and random inclination angles}

In an attempt to improve the low KS statistics seen in the $P$ and
$\dot{P}$ distributions for model 1, we search a range of different
luminosity indices.  Using Equation~\ref{eq:lum}, we vary $\alpha$
from $-1.9$ to $-0.8$ and $\beta$ from 0.1 to 0.9.  Our best result
occurs when $\alpha$ and $\beta$ take the values of $-1.0$ and 0.5, 
while FK06 found optimal values of $-1.5$ and 0.5.  As discussed in 
FK06, simulation techniques vary from model to model, and small 
deviations may occur between models in the search for optimal parameters.

So far we have assumed orthogonal rotators in the spin-down models.  A
more realistic and self-consistent approach is now followed by
assigning random inclination angles to all pulsars.

The results from model 2 are significantly better than model 1, as
seen in Table~\ref{tb:fom}.  At this point, we can see how much
improvement is gained by changing only two parameters, so we move on
to test other parameters to see if this improvement continues.

\subsubsection{Model 3: Beaming model and braking index variations}

Motivated by the improvements seen in model 2, we now relax the
requirement for the braking index $n=3$ and allow a range of braking
indices. Based on the observed sample, a sensible choice is to select
$n$ from a flat distribution in the range $2.5 < n < 3.0$.

Our previous models adopt the TM98 beaming model, which is used in
FK06.  However, we now consider computing the beaming fraction based
on a period-dependent opening angle law.

These results are worse than the previous model in terms of KS statistics, 
however, they are still better than the original model.  For model A, the 
birthrate decreases from nearly 4 pulsars per century to a more acceptable 
3 per century.  However, model B has just the opposite trend: the birthrate 
changes from 3 to 4 pulsars per century.

\subsubsection{Model 4: Inclination angle decay}

Our final model is motivated by the recent results of WJ08, who found
evidence for alignment of the spin and magnetic axes.  In both models
4A and 4B, the initially random inclination angles decay according to
Eq.~\ref{eq:chi} with an exponential decay timescale of
$t_d=10^7$~yr. Model 4A uses the spin-down law found in
Eq.~\ref{eq:pfk06c} while model 4B assumes the CS06 relation given in
Eq.~\ref{eq:pcs06}.  As shown in Table 2, both models perform very
poorly, with significantly lower figures of merit than the earlier
models without alignment.  For the case of model 4A, this poor
agreement can be understood when it is seen that Eq.~\ref{eq:pfk06c}
is essentially equivalent to an exponential decay of the magnetic
field --- more generally, an exponential decay of the braking
torque. From the work of FK06, and preliminary simulations we have
carried out, we know that a decaying braking torque is inconsistent
with the observations. For the case of model 4B, the strong dependence
on the spin-down model (Eq.~\ref{eq:pcs06}) with $\chi$ is also the
reason for the low figure of merit.

These results present a dilemma. On one hand, the empirical evidence
for alignment presented by WJ08 appears to be very robust. On the
other hand, we have found that magnetic dipole spin-down laws provide a
good description of the population without inclination angle decay.
The spin-down laws we have implemented depend critically on this
inclination angle. This implies that one way to resolve the
discrepancy is to modify the spin-down law so that inclination angles
are removed. While there is no physical basis for taking the $\sin^2
\chi$ term out of Eq.~\ref{eq:periodstart}, we find that a modified
version of model 4A which uses Eq.~\ref{eq:pfk06a} instead of
Eq.~\ref{eq:pfk06c} does produced improved results, though the overall
FOM (--6) is still lower than seen for the other models.
In a future paper, we intend to investigate this issue further.  Based
on our current results, however, it appears that the braking torque on
a pulsar is independent of its magnetic inclination angle.

Finally, for completeness, we have compared models 2A, 3A, and 4A with the 
pulse width model mentioned in \citet{slk+09}.  The birthrates for all 
three trials remain consistent with each model, however, the KS statistic
FOM is generally worse.  We conclude that changing the way in which pulse 
widths are modeled will not improve our results. 

\subsection{Individual parameters}

Separate from our optimal models, we test various pulsar parameters by
themselves in order to gain a better understanding of how they affect
the resulting pulsar population.  In this section, we
take a closer look at each of these parameters.

\subsubsection{Beaming}\label{sec:beaming}

We run multiple simulations to test the beaming models.  For each
trial, we alter only which beaming model is to be tested, and leave
all other parameters the same.  The constant beaming fraction model,
or one in which 20\% of all pulsars beam towards Earth, and the NV83
beaming model both give poor results.  The KS statistics using these
models are very close to zero for all simulations.

The beaming models of LM88, B90, and TM98 each give much better
results.  The TM98 model yields the best population of pulsars, but it
is not statistically better than the other two models.  Finally, we
try a period-dependent model, similar to one described in WJ08.  This
gives us our best result overall result in terms of KS statistics and
birthrates.

Based on these results, we can say a requirement for the population is a
period-dependent beaming model, as well as small opening angles,
similar to LM88, B90, and TM98.  These three beaming models, along
with WJ08, are all statistically about the same. 

\subsubsection{Braking Index}\label{sec:braking}

As a test of braking indices, we use our simulation Model 2A.  We try
indices ranging from 3.2 to 2.5, as well as a very realistic random
model in which indices are chosen at random between 2.5 and 3.0.  We
find very few differences when varying the braking indices, as shown
in Table \ref{tb:braking}.  Due to the lack of any significant
difference between various braking indices, we can adopt the realistic
random model for use in our simulations.

It is useful to again look back at the A models from
Table~\ref{tb:fom}.  We notice a decrease in birthrate when going from
model 2 to model 3.  The two parameters that are altered between the
two models are the braking index and the beaming model.  The results
presented in Table \ref{tb:braking} make it clear that this drop in
birthrate is not entirely due to the braking indices, and thus the
angle-dependent beaming model must affect the birth rate as well.

\begin{table} 
\caption{We show our results from 9 simulations with different values of the
pulsars' braking indices.  This is an implementation of the our Model 2A, 
changing only the braking index during each simulation.  The resulting birth 
rate, in units of pulsars born per century, and FOM (as described previously) 
is shown for comparison purposes.} 
\label{tb:braking}
\begin{center}
\begin{tabular}{ccc}
\hline
Braking Index & ${\cal R}$ (psrs/century) & FOM\\
\hline
$n=3.2$       & 3.75 & $-$3.78\\
$n=3.1$       & 3.99 & $-$4.09\\
$n=3.0$       & 3.56 & $-$1.75\\
$n=2.9$       & 3.51 & $-$1.63\\
$n=2.8$       & 3.52 & $-$2.27\\
$n=2.7$       & 3.53 & $-$3.66\\
$n=2.6$       & 3.25 & $-$1.41\\ 
$n=2.5$       & 3.28 & $-$2.74\\
$2.5<n<3.0$   & 3.60 & $-$1.38\\
\hline
\end{tabular}
\end{center}
\end{table}

\subsubsection{Luminosity}\label{sec:luminosity}

We run many luminosity simulations by changing the power law indices
of $P$ and $\dot{P}$.  The best result we obtain yields indices of
$-1.0$ for $P$ and 0.5 for $\dot{P}$.  Our trials also include a
simulation involving random luminosities, i.e.~no period dependence.
The results are less than ideal, and not close to the results we
obtain when using a period-dependent luminosity distribution.  Thus,
we conclude that there must be some period dependency for the
luminosity.  Similar conclusions were reached by FK06, however, their 
optimal value for the $P$ power law index was $-1.5$.

For our ideal simulations, our underlying luminosities have a
distribution that is very similar to the log-normal distribution found
in Fig.~15 of FK06.  In this regard, we note that this distribution
provides, in our opinion, the best current estimate of the pulsar
luminosity function. Unlike previous power-law models \citep[see,
e.g.,][]{lfl+06}, the log-normal does note require a lower luminosity
cutoff and is strongly recommended for studies which require some
reasonable assumption about the luminosity function.

\subsubsection{Covariances}\label{sec:covariances}

With such a large number of model parameters in our simulations, we
are mindful during our study to keep track of any interdependencies
that might be expected.  The biggest covariances we find involve the
luminosity.  Changing the braking index along with the luminosity
function can be done in such a way that two very similar results
can be obtained with two completely different luminosity
distributions and braking indices.  For example, $n$=3.0 and power
indices of $-1.5$ and 0.5 could yield the same results as $n$=2.7 and
indices of $-1.2$ and 0.8.  A similar relationship is found between
the luminosity and the alignment angle.  Running a simulation with all
angles equal to 90$^{\circ}$ might give similar results to another
simulation with all angles equal to 45$^{\circ}$ and having luminosity
indices of $-1.8$ and 0.9.  Most of these covariances are beyond the 
scope of this paper, simply due to the large matrix of parameters that 
would be analysed.

\subsubsection{Death line}

For completeness, we run simulations of both models that do not use a
death line.  As one might expect, we obtain an overabundance of high
period pulsars that have very low luminosities.  Our results here just
confirm that, within the framework of the models we have developed
here, a death line is indeed required when modeling a distribution of
pulsars.  While it might be possible to construct a model in future
which does not require a death line, we are not currently aware of any
straightforward means of achieving this.

\section{CONCLUSIONS}\label{sec:conclusions}

We have successfully implemented two pulsar spin-down models and
accounted for selection effects as far as possible to synthesize
populations of observable radio pulsars. From a statistical
comparisons of various models, we have gained a better understanding
of how their parameters affect the spin-down evolution of pulsars on a
$P$--$\dot{P}$ diagram.  Our main conclusions are as follows:
\begin{itemize}

\item The magnetic dipole spin-down model from FK06 works best with
our simulations.  We are able to reproduce the results from their
paper and obtain some improved results by modifying some of their
parameters. The model of CS06 produces poorer results. Further
modifications of this model appear to be required to improve it.

\item The braking index, $n$, does not have a significant impact on
our results.  Having a braking index of $n$=3.0 works just as well as
the rather unphysical scenario of $n>$3.0.  For our optimal
simulation, we use the most realistic model, which picks a random
braking index between 2.5 and 3.0 for each pulsar.

\item The optimal configuration for magnetic inclination angles
is a random distribution. Models in which the inclination angle
decays on timescales of $\sim 10^7$~yr do not reproduce the
observations well. Further modifications to the spin-down laws
appear to be required in order to account for the strong
empirical evidence for alignment found by WJ08.

\item Pulsar luminosities must have a period dependence.  We are able to
use a power law for the period and period derivative to replicate the
$P$--$\dot{P}$ diagram.  By tweaking the exponents in that law, we can
alter the resulting distribution of pulsars on our diagram. Even with
period dependent luminosity laws and beaming models, a death line is
required to explain the dearth of pulsars in the lower right of the
$P$--$\dot{P}$ diagram.

\end{itemize}
Studying these relationships between the $P$--$\dot{P}$ diagram and the various
pulsar parameters allowed us to become more aware of covariances, eliminate a
few parameter models, and overall obtain better insight to the behaviour of
pulsars as they evolve across the $P$--$\dot{P}$ diagram.  Future studies will
enable us to further narrow down some parameter models and, in particular,
allow us to investigate the issue of magnetic alignment.

\section*{Acknowledgements}

This work was supported by a West Virginia EPSCoR Research Challenge Grant 
awarded to the West Virginia University Center for Astrophysics.

\bibliographystyle{mn2e}
\bibliography{references}

\appendix
\section{An open-source approach to pulsar population synthesis}

Following the initial version described by \citet{lfl+06}, the source
code used in this work is freely available at {\tt
http://psrpop.sourceforge.net}.  We briefly describe the new features
of the pulsar simulation package, {\tt psrpop}, as well as announce
the availability of a website, {\tt http://psrpop.phys.wvu.edu}, used
to investigate past and future pulsar surveys.

The biggest new feature of the software package is the ability to
evolve a pulsar's spin period in time.  Currently, spin-down models
from FK06 and CS06 are available for use in the simulations.  By
having this evolution feature, we can now watch a pulsar's life cycle,
from birth to death, and observe how its period changes with time.
The program also has the ability to use various beaming models,
luminosity laws, and alignment angle functions.  These allow for
further studying of the individual pulsars, and a more complete
understanding of pulsar populations as a whole.

The {\tt psrpop} website currently has the capability of surveying any
of the eight model populations generated for this paper, and future
models will also be available for use. The user can ``search'' these
model populations using previous surveys such as the Parkes Multibeam Survey,
theoretical surveys such as one using the proposed Square Kilometer
Array, or they can create their own survey.  One of the main benefits
of running a custom survey is the ability to predict yields of future
surveys.

After surveying one of the pulsar populations, the website outputs the
total number of pulsars detected, a $P$--$\dot{P}$ diagram, and and
few comparison plots.  These plots contain histograms of the pulsar
properties detected in the survey that overlay histograms of the
observed pulsar population.  Some possible properties that can be
shown in the histogram plots are pulse period, period derivative,
dispersion measure, and flux density.
\end{document}